\documentclass{article}

\usepackage{arxiv}

\usepackage[utf8]{inputenc} 
\usepackage[T1]{fontenc}    
\usepackage{hyperref}       
\usepackage{url}            
\usepackage{booktabs}       
\usepackage{amsfonts}       
\usepackage{nicefrac}       
\usepackage{microtype}      
\usepackage{lipsum}		
\usepackage{graphicx}
\usepackage{natbib}
\usepackage{doi}
\usepackage{multirow}
\usepackage{subcaption}

\title{Epistral Network: Revolutionizing Media Curation and Consumption through Decentralization}

\author{ \href{https://orcid.org/0000-0001-5431-6367}{\includegraphics[scale=0.06]{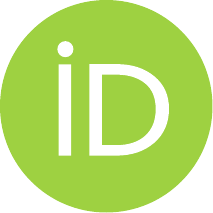}\hspace{1mm}Dipankar Sarkar} \\
  Epistral Protocol \\
  \texttt{me@dipankar.name} \\
	\AND
	Shubham Upadhyay \\
	Epistral Protocol \\
	\texttt{shubham@epistral.xyz} \\
}

\hypersetup{
pdftitle={Epistral Network: Revolutionizing Media Curation and Consumption through Decentralization},
pdfsubject={cs.CR},
pdfauthor={Dipankar Sarkar, Shubham Upadhyay},
pdfkeywords={Blockchain, Social Media,Representation},
}

\begin{document}
\maketitle

\begin{abstract}
Blockchain technology has revolutionized media consumption and distribution in the digital age, allowing creators, consumers, and regulators to participate in a decentralized, fair, and engaging media environment. Epistral, an innovative media network that leverages blockchain technology, aims to be the world's first anti-mimetic media curation and consumption network, addressing the core challenges facing today's digital media landscape: unfair treatment of creators and manipulative consumer algorithms, and the complex task of effective regulation. This paper delves into the conceptualization, design, and potential impact of epistral and explores how it embodies McLuhan's and Girard's theories within the realm of blockchain technology and draws from Hayden's critique of democratic representation. The paper analyzes the challenges and opportunities presented by this new network, providing a broader discourse on the future of media consumption, distribution, and regulation.
\end{abstract}


\section{Introduction}

In an era where digital media consumption is ubiquitous and constantly evolving, the profound observation of Marshall McLuhan, "the medium is the message," has never been more relevant. McLuhan's assertion, originating from his 1964 work "Understanding Media: The Extensions of Man," posits that the medium through which content is conveyed fundamentally shapes and controls the scale and form of human association and action \citep{mcluhan1964understanding}. This principle, while initially applied to the media landscapes of the mid-20th century, finds renewed significance in the context of today's digital networks, particularly when examining the intersection of media, technology, and society.

Drawing from McLuhan's foundational concepts, Girard's mimetic theory provides an alternative perspective on contemporary media dynamics \citep{girard1965deceit}. Girard posits that human desires are inherently imitative, originating from others, thus creating a cycle of imitation and competition that is prevalent in today's culture driven by social media. In the context of modern internet recommendation systems, there is a tendency for bias towards creating echo chambers for users. These systems are designed to present users with content similar to what they have previously encountered or engaged with, rather than offering diverse or challenging content. Consequently, users may be exposed to a limited range of ideas and opinions, reinforcing their existing beliefs and attitudes. This phenomenon can have detrimental effects on the cultivation of a well-informed and healthy society, as it can facilitate the dissemination of misinformation and the perpetuation of established power structures.

In the contemporary digital age, the intersection of these theories becomes increasingly relevant, especially when considering the challenges faced by content creators, consumers, and regulators within the digital media landscape. Content creators often grapple with unfair systems that exploit their work without equitable compensation, while consumers are subjected to aggressive algorithms designed to manipulate their attention and preferences. Regulators, on the other hand, struggle to balance the need for free expression with the necessity of maintaining societal norms and values in an ever-changing digital environment.

Grant M. Hayden, in "The False Promise of One Person, One Vote," examines the complexities of political equality in democratic systems, highlighting how individual influence is often misrepresented or diluted in larger democratic processes \citep{hayden2003false}. This discussion parallels the challenges in the digital media landscape, where individual voices and content creators are often overshadowed by larger entities or algorithms. Just as Hayden critiques the democratic ideal of equal representation, the landscape of digital media faces similar issues of unequal representation and influence.

The proposed solution to these challenges is Epistral, an innovative media network that leverages blockchain technology. Blockchain, first conceptualized by Satoshi Nakamoto in the creation of Bitcoin \citep{nakamoto2008bitcoin}, offers a decentralized and transparent platform that inherently challenges the traditional centralized models of content distribution and consumption. The Steem network, a blockchain-based social media platform, serves as a technical inspiration for Epistral due to its novel approach to content creation and distribution \citep{steemit2016steem}.

Epistral aims to be the world's first anti-mimetic media network, utilizing blockchain not just as a technology but as a medium to foster a new paradigm in content consumption and distribution. This approach seeks to address the core challenges facing today's digital media landscape: unfair treatment of creators, manipulative consumer algorithms, and the complex task of effective regulation. Unlike traditional platforms, Epistral's network does not focus on preventing intellectual property issues or serving as a store of value. Instead, it focuses on providing decentralized tools that empower content producers and facilitate a fair, engaging, and regulator-friendly media environment.

In this light, Hayden's insights provide a critical lens through which to assess the potential of blockchain in addressing the democratic deficits in digital media. We go beyond tackling the issues of how creators are treated and how algorithms manipulate content. Our vision is a future where successful regulation supports a flourishing media environment, while also maintaining the core principles of opposing censorship, promoting free speech, and embracing decentralization.

This paper delves into the conceptualization, design, and potential impact of Epistral, exploring how it embodies McLuhan's and Girard's theories within the realm of blockchain technology and draws from Hayden's critique of democratic representation. By analyzing the challenges and opportunities presented by this new network, we aim to contribute to the broader discourse on the future of media consumption and distribution in the digital age, as well as its implications for political equality and democratic processes.

\section{Background}

The advent of blockchain technology has catalyzed a wave of innovation in various sectors, notably in the realm of digital media. This section delves into the background and inspiration behind the creation of Epistral, highlighting key developments in blockchain-based media platforms that have paved the way for this novel network.

\begin{figure}[ht]
    \centering
    \includegraphics[width=0.80\linewidth]{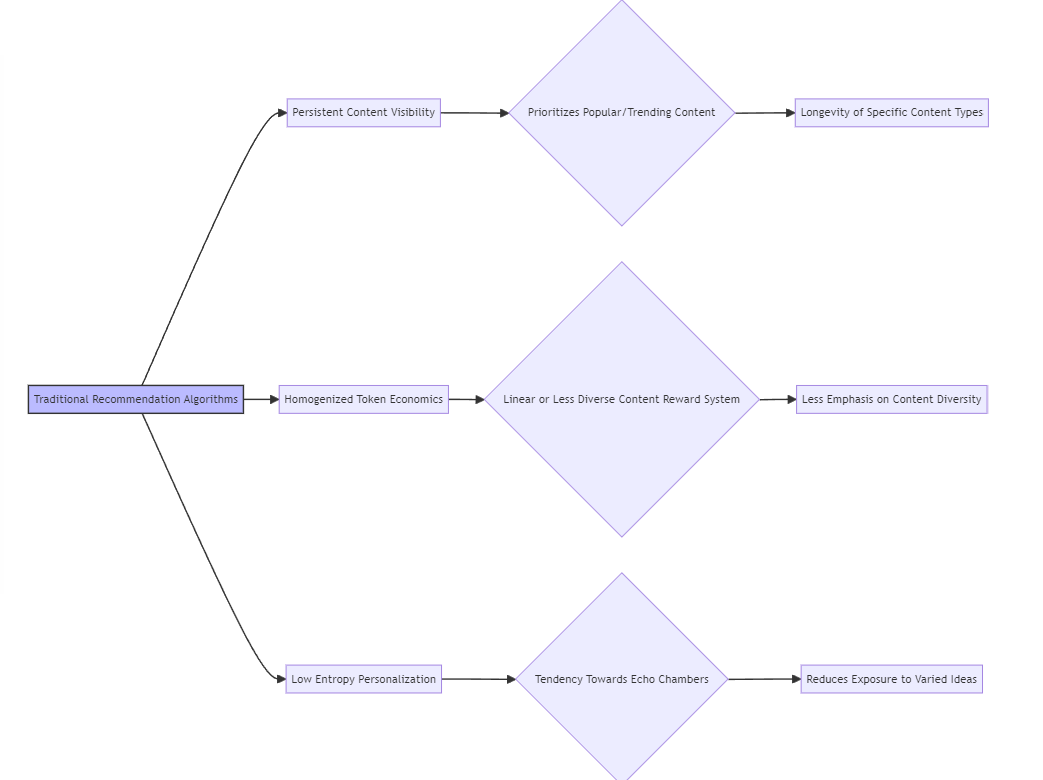}
    \caption{Challenges in recommendation systems}
    \label{fig:media-fair}
\end{figure}

\paragraph{Monetized social feeds \& Polarisation of Society} The popularity and widespread use of social media platforms such as Facebook, Instagram, and TikTok have grown significantly. These platforms offer users the ability to create and share content with a large audience. However, they also have the potential to create echo chambers, where users are only exposed to content that aligns with their initial preferences. This is particularly evident for users who are exposed to biased videos, as they are more likely to encounter similar content in the future. As a result, there has been a noticeable increase in the dissemination of misinformation videos, as users are more likely to be exposed to content that reinforces their existing beliefs. Furthermore, this limited exposure to diverse perspectives can lead to a lack of understanding and empathy towards different viewpoints, ultimately contributing to a more polarized society.

\paragraph{Emergence of Blockchain in Social Media} The integration of blockchain technology into social networks began as an effort to address the centralization and control prevalent on traditional platforms. The inherent characteristics of the blockchain of decentralization, transparency, and immutability offered a new paradigm for content creation, distribution, and monetization, fundamentally challenging existing media models.

A variety of blockchain-powered media outlets have been created, each of them providing distinctive perspectives and approaches that have been taken into account in the formation of Epistral: a Decentralized Platform for Content Aggregation and Incentivization.

D.tube, one of the pioneers in blockchain-based video platforms, emerged as a decentralized counterpart to YouTube. Initially, it was built on the Steem blockchain and utilized the InterPlanetary File System (IPFS) for decentralized storage. Users were rewarded with STEEM tokens, which demonstrates how the blockchain could incentivize content creation and engagement. Subsequently, D.tube moved to its own blockchain Avalon, while still on the HIVE network, and introduced the DTUBE coin, marking a significant evolution in its approach.

In the realm of livestreaming, Dlive began its journey on the Steem blockchain, but underwent several transitions. It moved to the Lino blockchain and later to TRON, illustrating the fluid nature of blockchain applications in media. Dlive's journey is indicative of the broader search for optimal operational frameworks within blockchain-based media platforms.

TravelFeed, originally a Steem ecosystem project, provided a curated blogging platform focused on travel content. It later transitioned to the Hive blockchain, allowing creators to use their own domains. This shift highlighted the ongoing evolution of blockchain platforms, particularly in enhancing content creators' autonomy and monetization opportunities.

3Speak, a platform dedicated to live streaming and video content, emphasized free speech. It was initially built on Steem \citep{steemit2016steem} and IPFS \citep{benet2014ipfs}. However, following Steem's acquisition by TRON, 3Speak, along with many other Steem-based decentralized applications (dApps), migrated to the Hive blockchain. This move highlighted the significance of community governance and blockchain's role in maintaining platform independence.

Finally, Dlike emerged as a decentralized blend of Medium and Instagram, offering a platform for sharing social media posts. It featured elements like trending content and topic filters, and its layout bore a resemblance to Steemit's. The existence of Dlike underlined the potential of blockchain technology to replicate and enhance existing social media functionalities.

These platforms collectively demonstrate the versatility and transformative potential of blockchain technology in various domains of online media, from video streaming to social blogging, each adapting and evolving in its unique way.

\paragraph{Epistral's Conceptual Foundation} The experiences of these pioneering platforms have been instrumental in shaping the vision of Epistral. The network draws on the successes and challenges faced by its predecessors, with the aim of refine and advance the application of blockchain in media. Epistral's unique approach is further inspired by the theories of McLuhan and Girard, integrating their insights into the design and functionality of the network. By combining the technical advancements demonstrated by platforms like D.tube, Dlive, TravelFeed, 3Speak, and Dlike with a deep understanding of media theory, Epistral positions itself as a groundbreaking entity in the blockchain media landscape.

In conclusion, Epistral is a product of the technological advances of blockchain in media and the theoretical frameworks of McLuhan and Girard. This combination is intended to tackle the current issues in media consumption, providing a decentralized, equitable, and stimulating platform for creators, consumers, and regulators.

\section{The Epistral Network: Vision and Goals}

The Epistral Network is a revolutionary response to the multifarious issues and trends of the current digital media environment. Its purpose is to rethink and redefine the way content is produced, distributed, and managed, taking into account both the theoretical foundations of media experts and the technical progress in blockchain technology. 

At its core, Epistral Network is an algorithm company that focuses on promoting credible, diverse, and enjoyable platform recommendations. The credibility of these recommendations is determined by a combination of reputation and personal investment, with each user interaction carrying a unique value. Although the network is open to everyone, it effectively addresses the issues of content discovery and incentivization by providing a solution that caters to credible and enjoyable recommendations.

The vision of Epistral is to create a media network that surpasses traditional boundaries and restrictions, creating an ecosystem where fairness, diversity, and autonomy are not just ideals, but actual realities. The fundamental objectives of the Epistral Network are multifaceted, each tackling a key element of the media system:

\begin{enumerate}
    \item \textbf{Diversity in Media Consumption}: Epistral aims to break the echo chambers and algorithmic biases prevalent in existing media platforms. The network is designed to promote a diverse range of content, encouraging exposure to varied perspectives and ideas. This goal is not only about offering a wide array of content but also about ensuring that this content is accessible and visible to a broad audience, thus enriching the overall media consumption experience.
    
    \item \textbf{Fairness in Content Distribution}: Central to Epistral's mission is the establishment of a fair and equitable system for content creators. This goal addresses the imbalance and exploitation often experienced in traditional media platforms, where creators' contributions are undervalued. By leveraging blockchain technology, Epistral ensures that content creators are justly rewarded for their work, fostering an environment where creativity and originality are incentivized.

    \item \textbf{Sovereignty Among Users and Regulators}: Sovereignty in the Epistral context refers to the empowerment of both users and regulators within the media ecosystem. For users, this means having control over their data and the content they consume, free from manipulative algorithms and invasive advertising practices. For regulators, Epistral offers a transparent and accountable framework, enabling them to enforce regulations effectively while respecting the principles of free expression and privacy.
\end{enumerate}

Epistral Network is revolutionizing the media industry by combining media expertise with blockchain technology. This innovative approach is ushering in a new era of digital media that prioritizes fairness, diversity, and autonomy, while also fostering genuine creativity and engagement. By adhering to these principles, the network not only addresses current challenges in digital media but also paves the way for a more inclusive, dynamic, and fair media landscape in the future. 

\section{Core System Architecture: Attribution and Governance}

The architectural design of the Epistral Network is a sophisticated amalgamation of various blockchain technologies and principles, thoughtfully orchestrated to address the nuanced needs of creators, consumers, and regulators in a Hive-based blockchain environment. This multidimensional architecture ensures that each participant group is not only integrated but also derives optimum benefit from the network's diverse functionalities.

\begin{figure}[ht]
\centering
\includegraphics[width=1.0\linewidth]{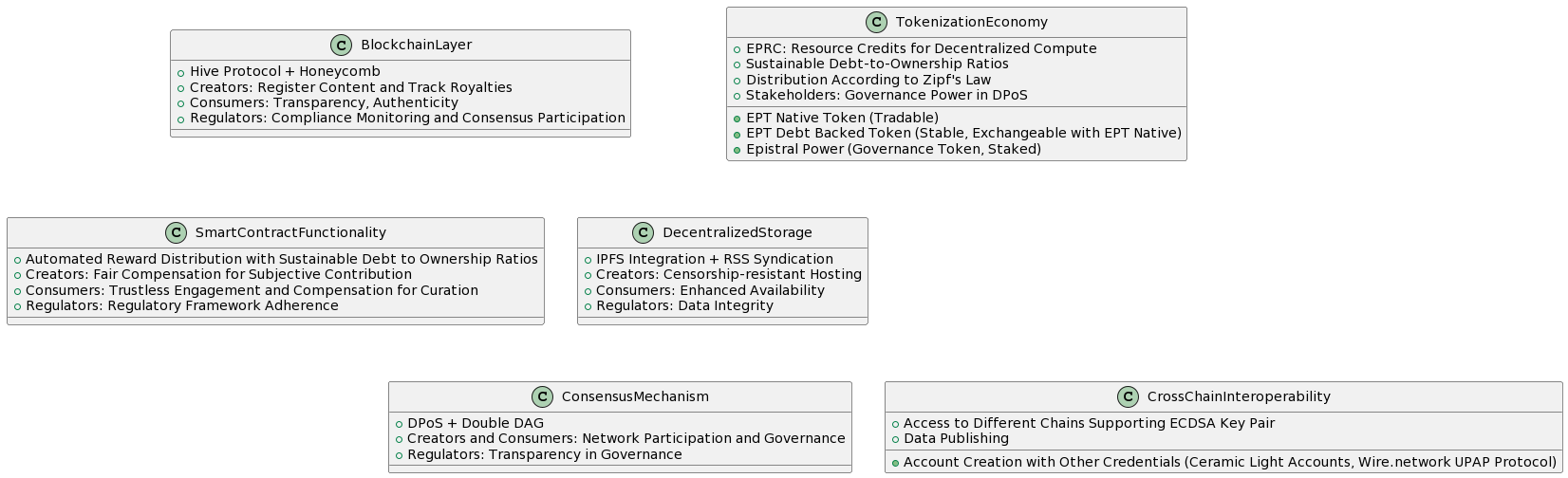}
\caption{In-depth System Components of the Epistral Network}
\label{fig:in-depth-system-components}
\end{figure}

The primary component of this ecosystem is the \textbf{Blockchain Layer}, which includes the Hive Protocol and Honeycomb. This layer plays a crucial role in creating a strong base for different parties involved. Content creators can use this platform to register their work and keep track of royalties, taking advantage of the unchangeable nature of blockchain that permanently records their rights and contributions. Consumers benefit from increased transparency and authenticity in their interactions with content, as evidenced by secure and unalterable transaction records. Regulators find this layer essential for monitoring compliance and promoting consensus participation, thus aligning with legal and regulatory requirements.

The Epistral platform incorporates a \textbf{Consensus Mechanism} that combines DPoS with a Double DAG system. This mechanism aims to establish a democratic governance model where both creators and consumers actively engage in network governance. By doing so, they contribute to a decision-making process that is transparent and inclusive. Additionally, this mechanism provides regulators with a transparent view of governance and network operations.

The \textbf{Smart Contract Functionality} in Epistral stands as a beacon of innovation. These contracts are engineered to distribute rewards automatically, maintaining sustainable debt-to-ownership ratios. Creators are fairly compensated for their subjective contributions, fostering a sense of equity and recognition. Consumers, on the other hand, engage in a trustless environment, rewarded for their curation efforts. This setup also ensures strict adherence to regulatory frameworks, automating compliance, and enhancing transparency.

The incorporation of native and debt-backed EPT tokens is a crucial aspect of Epistral's economic model, known as tokenization. Tokenization refers to the process of creating tokens within the network. The native EPT token is tradable, while the EPT debt-backed token is designed to provide stability. These tokens operate within sustainable debt-to-ownership ratios. Epistral also utilizes the Epistral Power token for governance, which is staked by participants. Additionally, the use of EPRC tokens for decentralized compute resources demonstrates a comprehensive tokenization strategy. This strategy ensures a fair and balanced distribution of tokens among stakeholders, in accordance with Zipf's law \citep{zipf2013psycho}

The network's \textbf{Decentralized Storage} is a vital component, which is accomplished by integrating IPFS and utilizing RSS Syndication. This characteristic provides creators with a strong and resistant hosting platform that is not subject to censorship, ensuring the durability and reliability of their content. Additionally, users benefit from improved content accessibility, free from the risks associated with centralized hosting. Regulators also appreciate this decentralized approach as it guarantees the trustworthiness and genuineness of data, which is essential for maintaining overall confidence in the system.

\textbf{Cross-Chain Interoperability} is another notable feature of the Epistral Network. It facilitates the creation of accounts using other credentials, possibly through a naming service like Ceramic light accounts or Wire.network's UPAP protocol. This interoperability allows access to different chains supporting ECDSA key pairs, thereby enhancing the network's reach and utility. It broadens the spectrum for content distribution, allowing creators and consumers to interact with a wider array of blockchain ecosystems.

The architecture of the Epistral Network is a prime example of innovation, as it combines the Hive protocol with advanced features that are specifically designed to meet the diverse requirements of its various stakeholders. This complex design not only aligns with the network's strategic vision, but also represents a significant advancement in the development of digital media platforms based on blockchain technology. Epistral aims to cater to the unique needs of creators, consumers, and regulators, positioning itself as a versatile and forward-looking platform in the field of blockchain.

\section{Recommendation system: Diversity and Fairness}

The Epistral Network is striving to revolutionize digital media consumption and creation by introducing anti-mimetic principles to counter the widespread inclination towards mimetic desire and echo chambers in social media. This section will explore the methods used by Epistral to guarantee content variety, impartiality, and a break from mimetic tendencies.

The implementation of antimimetic principles in the Epistral Network is a purposeful and creative attempt to combat the homogenizing effects of traditional social media algorithms and predict the authenticity of subjective proof of work using cumulative reputation of stakeholders. Through the limited lifespan of content, graph-based token production, and a recommendation algorithm emphasizing embedding entropy, Epistral encourages a varied, dynamic, and equitable media environment.

These mechanisms not only enhance the user experience by exposing them to a wider array of content but also ensure that the network's economy thrives on the principles of diversity and originality. By doing so, Epistral sets a new standard for digital media platforms, one that champions diversity and combats the mimetic cycles prevalent in today's digital landscape.

\subsection{Content Lifecycle}
The Epistral network has established a specific timeframe for the lifecycle of its content, which is determined by network governance. This intentional decision is made to encourage a constant stream of new and varied content. The distribution of tokens is based on the level of engagement that the content receives during this timeframe. This strategy encourages creators to consistently generate engaging and thought-provoking content, thus reducing the risk of content becoming stagnant and the formation of echo chambers.
\begin{figure}[ht]
    \centering
    \includegraphics[width=0.80\linewidth]{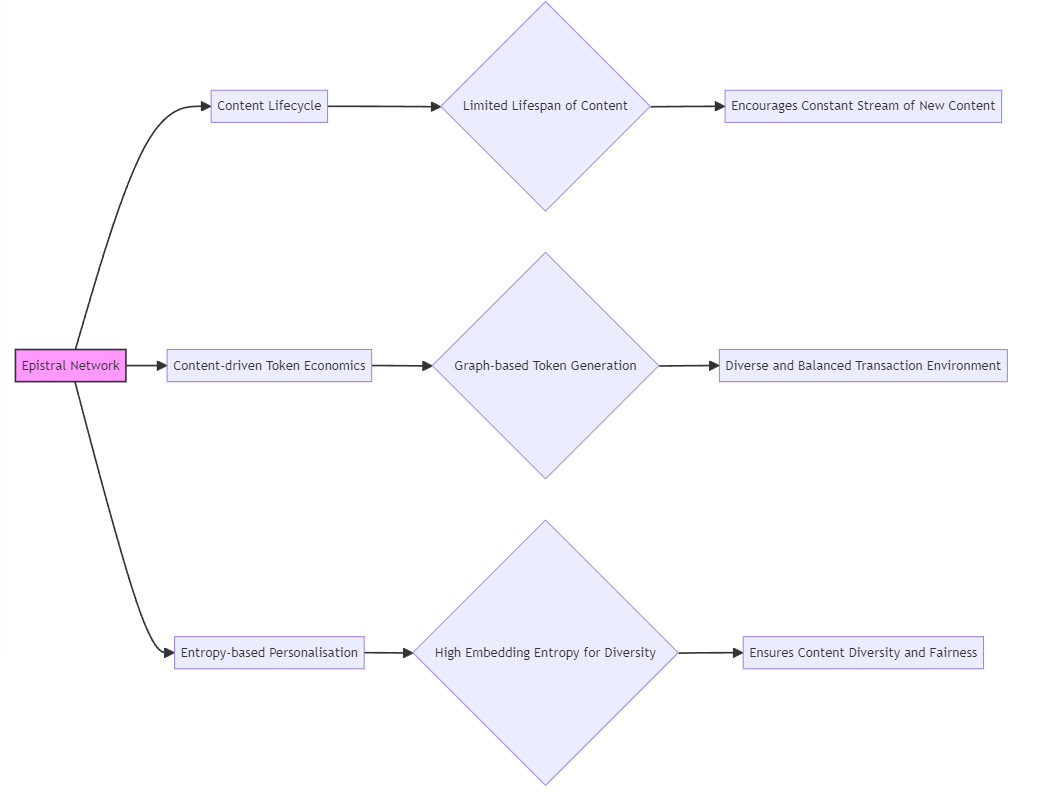}
    \caption{Media fairness, diversity and transparency}
    \label{fig:media-fair}
\end{figure}

\subsection{Content-driven Token Economics}
Epistral utilizes a graph-based approach to token generation, which evaluates the distribution of content across its blockchain. This method assesses how content is shared and interacted with throughout the network, focusing on diversity and reach. The algorithm takes into account both the amount of content and the type of transactions within the network. A higher variety of content and a balanced transaction environment lead to increased token generation.

This system encourages a vibrant and varied content landscape, with the Epistral token economy being directly linked to the diversity of content and its usage. This connection ensures that economic incentives are in line with the network's anti-mimetic goals, rewarding content that contributes to a diverse and dynamic media ecosystem.

\subsection{Entropy-based Personalisation}
Epistral's personalized recommendation algorithm is specifically designed to ensure that the content consumed by users maintains a high level of embedding entropy. This concept refers to the diversity and unpredictability of the content recommendation system. While the network can establish target entropy levels, the primary objective is to optimize for diversity and fairness in the content feed. This approach helps to prevent users from being exposed to a limited range of content and reduces the risk of algorithmic echo chambers. 

Unlike traditional media platforms that often trap consumers in localized content loops, Epistral's algorithm actively opposes this trend. By prioritizing content diversity and embedding entropy, the platform ensures that users are not confined to narrow content silos but are instead exposed to a wide variety of ideas and perspectives.

\textit{The recommendation system does not give equal opportunities to all content}. If a piece of content is of high quality, it will have a higher chance of being surfaced due to the system's design. We do not use any anti-bot system, as the entropy-based system prevents any specific narrative from dominating the consumer feed. Regardless of how something is curated, its value only increases in relation to its surrounding content, not globally. 

For example, if there are 100,000+ bot-produced articles about "Elon Musk" with high votes, but only 100 articles about Albert Einstein, the "Elon Musk" articles will not take over the user's feed, no matter how much curation is done. Instead, a few good articles in the "Elon Musk" neighborhood (in terms of semantic similarity) will be shown more frequently compared to others in that neighborhood.

\section{Token Economy: Incentives and Rewards}

The Epistral Network introduces a sophisticated token economy centered around its native cryptocurrency, the Epistral Token (EPT). This economy is designed to incentivize and reward the various activities within the network, aligning the interests of creators, consumers, curators, and other stakeholders. The token economy is fundamental to achieving the network's goals of fairness, diversity, and sovereignty in media consumption and creation.

\begin{figure}[ht]
    \centering
    \includegraphics[width=1.0\linewidth]{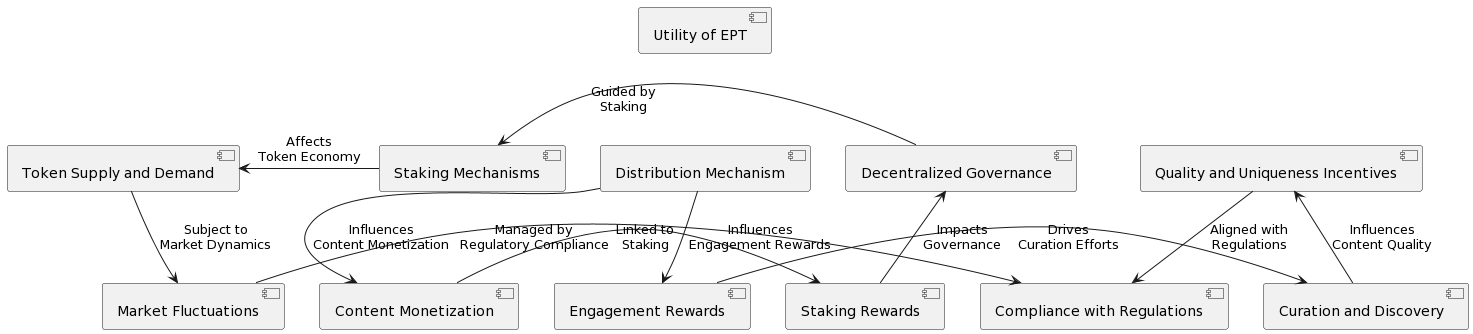}
    \caption{Token economy}
    \label{fig:token-economy}
\end{figure}

The EPT token is a key component of the Epistral Network's blockchain-based media platform, helping to create a dynamic and interactive environment. By utilizing the token, Epistral ensures that the interests of all stakeholders are aligned, encourages the production of high-quality content, and gives users the opportunity to take part in the network's governance. This token economy not only provides a fair and rewarding system for digital media, but also demonstrates the innovative power of blockchain technology in developing sustainable and equitable economic models in the digital world.

\paragraph{Token Functionality and Distribution}
The Epistral ecosystem utilizes the EPT token for multiple purposes: as a medium of exchange, a measure of value, distribution of compute, management of liquidity and a tool for governance. Users can earn EPT through activities such as content creation, curation, and participation in network governance. Epistral has a nuanced distribution mechanism that rewards users based on their contribution to the network, including content creation and curation effectiveness, community engagement, and other valuable activities. The distribution is managed by smart contracts ensuring sustainable reducing inflation, transparency, and fairness.

The initial distribution of tokens consists of both liquid and staked tokens, with the distribution changing based on the debt:ownership ratios. However, users always have the opportunity to acquire a proportional ownership stake in the network. They can do this by either purchasing tokens through capital contributions or by earning them through their work. It is worth noting that reputation cannot be obtained solely through capital contributions.

\paragraph{Incentives for Creators}
Creators in the Epistral network are incentivized through direct monetization of their content. They earn EPT tokens based on community engagement metrics like views, likes, and shares, promoting the creation of high-quality, original content. Additionally, creators can stake their EPT tokens to earn extra rewards and gain more influence in network governance, allowing them to participate in important platform decisions.

\paragraph{Incentives for Consumers}
Consumers are rewarded with EPT tokens for engaging with content, such as viewing, liking, commenting, and sharing. This model fosters active participation and enhances the network's vibrancy. Effective curation and discovery of trending or valuable content by consumers are also rewarded, encouraging a community-driven approach to content promotion.

\paragraph{Governance and Staking}
The Epistral network's governance is decentralized, with token holders having a say in its operation. The more EPT a user stakes, the greater their voting power in decisions like feature developments and policy changes. Staking mechanisms not only enable participation in governance but also secure the network, with stakers receiving additional token rewards, aligning their interests with the network's success.

\paragraph{Token Economy Dynamics}
The supply and demand of EPT are regulated to maintain its value and utility within the ecosystem. Demand is driven by its various uses, including content monetization, access to premium features, and governance participation. The token economy is designed to reward the quality and uniqueness of content, encouraging creators to produce innovative and engaging material, thus aligning with Epistral's anti-mimetic ethos.

\paragraph{Regulatory Compliance}
Managing market fluctuations is critical to the stability of the EPT token. Epistral implements mechanisms to mitigate extreme volatility. Additionally, the token economy is designed with an acute awareness of regulatory requirements, ensuring compliance with financial laws and guidelines across various jurisdictions.
\section{Key Innovations: Consumer, Privacy and Compliance}

The Epistral Network, developed on the Hive blockchain protocol, presents a range of innovative features and functionalities designed to revolutionize the digital media landscape. These key aspects are tailored to meet the diverse needs of creators, consumers, and regulators, ensuring an equitable, engaging, and compliant media ecosystem.

\subsection{Consumer Engagement and Interaction}
For consumers, Epistral provides a personalized content discovery experience using advanced algorithms. This feature tailors the content exposure to each user's preferences and interests, thereby enhancing user engagement. The platform also incorporates interactive features, allowing consumers to engage with content through likes, comments, and shares. These interactions not only improve the user experience but also increase the content's visibility and the creator's potential for earning rewards.

Incorporating the anti-mimetic philosophy, inspired by the works of McLuhan and Girard, Epistral seeks to cultivate an environment where content is not merely a reflection of existing desires and trends but a platform for genuine expression and innovation. This approach challenges the conventional dynamics of content creation and consumption, encouraging users to engage with content that inspires original thought and creativity, rather than perpetuating cycles of imitation.

\subsection{Scalable and Secure Technology Stack}
To enhance its functionality and reach, Epistral supports cross-chain interoperability, facilitating seamless interactions with other blockchain networks. This interoperability expands the range of content and token exchange possibilities. Furthermore, the platform integrates oracle services to import external data onto the blockchain. These services provide valuable insights for content strategy and market trends, benefiting both creators and consumers.

Epistral places a high priority on security and privacy. Leveraging Hive's robust security infrastructure, the platform ensures the protection of user data and transactions. Advanced encryption and privacy-enhancing technologies are employed to safeguard user information and the integrity of content. Moreover, users are given control over their personal data, with options to manage privacy settings and data sharing preferences. This approach is in line with global privacy standards and caters to user expectations for data sovereignty.

In terms of scalability and performance optimization, Epistral utilizes Hive's DPoS consensus mechanism to achieve efficient transaction processing. This capability is essential for managing the large volumes of user interactions and content uploads. The network's architecture is specifically designed for scalability, ensuring that as the platform grows, it continues to provide a seamless and responsive experience for all users.

\subsection{Regulatory Compliance and Governance}
Epistral is committed to regulatory compliance and governance. The platform incorporates a transparent regulatory framework to ensure adherence to various legal standards, including content moderation, intellectual property rights protection, and regional content regulations. Additionally, Epistral employs a decentralized governance model, enabling stakeholders to propose and vote on changes. This approach ensures that the network evolves in a manner that aligns with the collective interests of its participants.

The Epistral Network's key features and functionalities represent a comprehensive approach to creating a decentralized, user-centric, and regulator-friendly media platform. By addressing the needs of creators, consumers, and regulators through innovative blockchain applications, Epistral stands as a pioneering solution in the digital media domain, poised to reshape how content is created, distributed, and consumed in the blockchain era.

\section{Challenges and Limitations}

The Epistral Network, a pioneering endeavor, is confronted with a variety of difficulties and restrictions that are inherent in its design and the wider environment in which it operates. To address these issues, a combination of technical advancement, economic stability, strategic planning, and compliance with regulatory requirements is necessary. It is essential for Epistral to overcome these obstacles in order to realize its ambition of transforming the digital media landscape through blockchain technology and anti-mimetic principles. As the platform progresses, ongoing assessment and adjustment will be essential for overcoming these challenges and ensuring long-term success.

\paragraph{Technical}
Epistral, like many other blockchain-based systems, is confronted with the major issue of scalability. It is essential for the network to be able to process a large number of transactions and interactions quickly and without any loss of speed or performance. Although Hive’s infrastructure offers scalability advantages, Epistral must continue to develop in order to accommodate the growing user activity and content volume. Another significant technical challenge is the complexity of implementing and maintaining advanced anti-mimetic algorithms. These algorithms, which are essential for guaranteeing a wide range of content distribution and tailored recommendations, must be able to adjust to changing user behaviors and content trends, necessitating ongoing development and optimization.

\paragraph{Economic and Tokenization}
In the realm of economics and tokenization, the volatility of the cryptocurrency market poses a considerable challenge. Fluctuations in the value of the EPT token can significantly impact the network's economy, affecting both creator incentives and user engagement. Managing this volatility to ensure the stability of the token is a major task. Additionally, designing a token economy that balances the incentives for creators, consumers, and curators fairly is a delicate task. Over-incentivizing one group could lead to imbalances and unintended network behaviors.

\paragraph{Content and User Engagement}
Ensuring consistent content quality is a complex issue for Epistral. Although the network encourages high-quality content creation, relying on community curation and engagement metrics for content valuation could result in quality disparities. Another major challenge is user adoption and retention. 

\paragraph{Regulatory and Compliance}
Navigating the complex network of regulations in different jurisdictions is a significant challenge for Epistral, operating in a global landscape. The platform must ensure compliance with various legal standards, especially in areas such as data protection, content regulation, and financial transactions. Furthermore, it is complex to balance the decentralized nature of blockchain with regulatory compliance. Decentralization offers numerous benefits, but can complicate the adhesion to certain regulatory frameworks that require centralized control or oversight.

\paragraph{Limitations}
Technological barriers present a limitation for less tech-savvy users engaging with Epistral. Simplifying the user experience and educating users about blockchain and cryptocurrencies are crucial steps to overcome this barrier. Additionally, as a decentralized platform, Epistral has limited control over user behavior and content. This lack of control can pose challenges in content moderation and in ensuring a safe and respectful online community.

\section{Case Studies and Applications}

This section aims to illustrate the practical implications and possibilities of the Epistral Network. Through hypothetical case studies and applications, we will demonstrate how the unique features of Epistral can tackle current issues in the digital media sphere and open up new possibilities for creators, consumers, and regulators.

\subsection*{Case Study 1: Empowering Independent Creators}
\begin{itemize}
    \item \textbf{Background}: Jane, an independent documentary filmmaker, struggles to find a platform that fairly compensates her for her work and offers her control over content distribution.
    \item \textbf{Application of Epistral}: On Epistral, Jane uploads her documentary. The platform's blockchain-based system ensures her ownership rights are protected. Her work gains traction due to the network's anti-mimetic content recommendation algorithm, exposing her documentary to a diverse audience.
    \item \textbf{Outcome}: Jane receives fair compensation in EPT tokens based on genuine audience engagement within the 15-day content lifespan. She reinvests some tokens to stake in the network, gaining a say in future governance decisions.
\end{itemize}

\subsection*{Case Study 2: Enhancing Consumer Experience}
\begin{itemize}
    \item \textbf{Background}: Alex, a content consumer, is tired of the repetitive and echo-chamber nature of traditional social media feeds.
    \item \textbf{Application of Epistral}: Alex discovers Epistral and experiences a content feed curated based on embedding entropy, exposing him to a variety of content genres and viewpoints.
    \item \textbf{Outcome}: Alex finds the content on Epistral more enriching and less biased. He actively participates in content curation, earning EPT tokens, which further enhances his engagement with the platform.
\end{itemize}

\subsection*{Case Study 3: Regulatory Compliance and Transparency}
\begin{itemize}
    \item \textbf{Background}: A regulatory body seeks to ensure compliance and transparency in digital content without imposing excessive controls.
    \item \textbf{Application of Epistral}: The body utilizes Epistral's transparent blockchain ledger and regulatory dashboards to monitor content and transactions, ensuring compliance with regional standards.
    \item \textbf{Outcome}: The regulator effectively oversees the digital content landscape on Epistral, ensuring it aligns with legal standards while appreciating the platform’s decentralized governance model.
\end{itemize}

\subsection*{Case Study 4: Cross-Cultural Content Collaboration}
\begin{itemize}
    \item \textbf{Background}: A global collaboration project aims to bring together content creators from diverse cultures to create a series of interconnected digital art pieces.
    \item \textbf{Application of Epistral}: Creators use Epistral to upload and link their content, benefiting from the platform’s cross-chain functionality and decentralized storage. The diverse content ecosystem of Epistral facilitates wider exposure.
    \item \textbf{Outcome}: The project achieves global recognition, with contributors receiving equitable compensation and recognition for their work. The initiative showcases the power of blockchain in fostering cross-cultural collaboration.
\end{itemize}

\subsection*{Case Study 5: Combatting Digital Content Monopolies}
\begin{itemize}
    \item \textbf{Background}: Small content creators face challenges competing against large digital content monopolies.
    \item \textbf{Application of Epistral}: Utilizing Epistral’s equitable token distribution and anti-mimetic recommendation engine, these creators gain a level playing field where quality and originality are key.
    \item \textbf{Outcome}: A more diverse range of creators emerge, breaking the dominance of larger content monopolies and enriching the digital content landscape with unique and varied perspectives.
\end{itemize}

\subsection*{Case Study 6: Transforming E-commerce Reviews through an Anti-Mimetic Approach}
\begin{itemize}
    \item \textbf{Background}: Emily owns a small e-commerce business that struggles to gain visibility and credibility due to the dominance of larger retailers and the prevalence of inauthentic reviews on traditional platforms.
    \item \textbf{Application of Epistral}: Emily chooses to utilize the Epistral Network to conduct her product evaluations. Epistral incorporates the reviews into the blockchain, guaranteeing their credibility and openness. The platform's anti-mimetic strategy ensures that the reviews are varied and authentic, capturing a broad spectrum of customer encounters.
    \item \textbf{Outcome}: The anti-mimetic algorithm effectively identifies different genuine reviews, preventing the problem of biased or uniform feedback. This openness and variety in reviews appeal to a larger audience for Emily's business, improving the credibility and reliability of her brand. Customers are more assured in their buying choices, as they know the reviews are authentic and reflect a diverse range of opinions, not just the most popular or imitative perspectives.
\end{itemize}

These case studies exemplify the transformative potential of the Epistral Network in various aspects of digital media creation, consumption, and regulation. By leveraging blockchain technology and anti-mimetic principles, Epistral addresses key challenges faced in the current media ecosystem, offering innovative solutions that benefit all stakeholders. As the platform evolves, it is poised to inspire new applications and case studies, further demonstrating its impact on the digital media world.

\section{Future Directions}

The Epistral Network is committed to staying at the forefront of blockchain technology. To enhance scalability, security, and user experience, the network plans to continuously integrate the latest advancements in blockchain technology. This includes exploring layer-2 solutions, sidechains, and other technological innovations that could significantly augment network performance.

In the realm of content moderation and enhancement, Epistral is turning towards \textit{artificial intelligence} as a key tool for the future. AI will play a crucial role in maintaining content quality, detecting and addressing harmful content, and optimizing the anti-mimetic content recommendation algorithms. The use of AI is expected to streamline these processes, ensuring a safer and more engaging platform for users.

Epistral is also focusing on expanding its \textit{cross-chain functionalities}. By enhancing its ability to interact seamlessly with a broader range of blockchain ecosystems, the network aims to increase its content reach, diversify its token economy, and facilitate wider user adoption. This expansion is seen as crucial for the network’s growth and integration within the broader blockchain community.

Another significant area of focus for the future is\textit{ user education and community building}. Epistral recognizes the importance of educating users about blockchain technology and the specifics of its platform. Building a strong, engaged community is seen as essential for the network's growth and sustainability. This approach is expected to foster a more informed and active user base, contributing to the overall health and dynamism of the platform.

Lastly,\textit{ regulatory collaboration and compliance} are top priorities for Epistral. The network aims to work closely with regulatory bodies to ensure compliance and adapt to the evolving legal landscapes. Epistral's goal is to be at the forefront of defining regulatory standards for blockchain-based media platforms, setting a benchmark for responsible and compliant operation in this emerging field.

\section{Conclusion}
The Epistral Network represents a groundbreaking contribution to the digital media landscape. Its innovative use of blockchain technology, combined with anti-mimetic principles, sets it apart from traditional media platforms. Key contributions of the Epistral Network include:
\begin{itemize}
    \item \textbf{Revolutionizing Content Monetization}: Epistral's token economy and smart contract-based monetization mechanisms offer a fair and transparent way for content creators to earn from their work, addressing a major gap in the current digital content market.

    \item \textbf{Promoting Content Diversity and Quality}: The network’s anti-mimetic approach to content recommendation and curation fosters a diverse and high-quality content ecosystem. This is a significant departure from the echo chambers and mimetic cycles prevalent in existing social media platforms.

    \item \textbf{Empowering Users and Decentralizing Governance}: By giving users a stake in network governance and rewarding participation, Epistral empowers its community and promotes decentralized decision-making. This model is pivotal in shaping a media platform that is truly by and for its users.

    \item \textbf{Enhancing Transparency and Regulatory Compliance}: The transparent nature of blockchain technology, as implemented in Epistral, provides a new level of transparency and accountability in digital media. This is a major step forward in balancing freedom of expression with regulatory compliance.

    \item \textbf{Setting a Precedent in Blockchain Media Platforms}: Epistral sets a precedent for future blockchain-based media platforms, showcasing the potential of this technology to transform how digital content is created, distributed, and consumed.
\end{itemize}

In conclusion, the Epistral Network not only addresses current challenges in the digital media sphere but also opens up new possibilities for fair, diverse, and engaging content consumption and creation. As the network continues to evolve, it stands poised to make a lasting impact on the digital media landscape, paving the way for more equitable, innovative, and user-centric media platforms.

\bibliographystyle{unsrtnat}
\bibliography{references}
\end{document}